\documentclass[aps,prl,groupedaddress,showpacs,showkeys]{revtex4} %,twocolumn

\def\tfrac#1#2{{\textstyle{#1\over#2}}}
\def\half{\tfrac{1}{2}}
\def\dg{\dagger}
\def\bra#1{\langle #1\vert}
\def\ket#1{\vert #1\rangle}
\def\redbra#1{( #1\vert\vert}
\def\redket#1{\vert\vert #1)}
\def\cd#1{c^\dg_{#1}}
\def\ct#1{\tilde c^{ }_{#1}}
\usepackage{graphicx}
\begin{document}

% Use the \preprint command to place your local institutional report
% number in the upper righthand corner of the title page in preprint mode.
% Multiple \preprint commands are allowed.
% Use the 'preprintnumbers' class option to override journal defaults
% to display numbers if necessary
%\preprint{}

%Title of paper
\title{Efficient matrix-vector products for large-scale
  nuclear Shell-Model calculations}

% repeat the \author .. \affiliation  etc. as needed
% \email, \thanks, \homepage, \altaffiliation all apply to the current
% author. Explanatory text should go in the []'s, actual e-mail
% address or url should go in the {}'s for \email and \homepage.
% Please use the appropriate macro foreach each type of information

% \affiliation command applies to all authors since the last
% \affiliation command. The \affiliation command should follow the
% other information
%\affiliation can be followed by \email, \homepage, \thanks as well.
\author{J. Toivanen}
%\homepage[]{Your web page}
%\thanks{}
%\altaffiliation{}

\affiliation{Department of Physics, University of Jyv\"askyl\"a, P.O. box 35,
FIN-40014, Finland}
%\email[]{jussi.toivanen@phys.jyu.fi}

\date{\today}

\begin{abstract}
  A method to accelerate the matrix-vector products of j-scheme
  nuclear Shell-Model Configuration Interaction (SMCI) calculations is
  presented. The method takes advantage of the matrix product form of
  the j-scheme proton-neutron Hamiltonian matrix. It is shown that the
  method can speed up unrestricted large-scale pf-shell calculations
  by up to two orders of magnitude compared to previously existing
  related j-scheme method. The new method allows unrestricted SMCI
  calculations up to j-scheme dimension $10^{10}$ to be made in more
  complex model spaces.
\end{abstract}

\maketitle

%\section{I. INTRODUCTION}

{\it Introduction:} The nuclear Shell-Model is the most general
microscopic nuclear model and is in principle able to describe all
properties of nuclei.  Nuclear Shell-Model Configuration Interaction
(SMCI) calculations in large and realistic single-particle (s.p.)
model spaces are, however, very difficult to make due to the extremely
large Hilbert space dimensions involved.  The continuous increase in
computing power has made it possible to make progressively larger
nuclear SMCI calculations in restricted model spaces.  Currently
existing nuclear SMCI methods/programs make it possible to calculate
nuclear wavefunctions exactly in the model spaces $sd$ and
$pf$~\cite{Caurier99} and in $pf_{5/2}g_{9/2}$ model space with
somewhat truncated calculations. In larger model spaces drastic
truncations (i.e. selection of allowed particle configurations) have
to be made. Two basic problems arise in large-scale SMCI calculations:
Large dimensions require huge amounts of storage space for the
calculated states, and the number of non-zero Hamiltonian matrix
elements becomes prohibitively large ($10^{14}-10^{16}$) even though
the Hamiltonian matrix stays very sparse. Therefore both available
memory and computational speed often become inadequate for the task at
hand.

Using very large computing resources to solve SMCI problems is the
brute force approach to circumvent these problems.  Alternatives to it
are to develop mathematical methods to truncate large calculations and
to keep only physically relevant degrees of freedom in the
wavefunctions, or to make the most time consuming parts of SMCI
calculations more efficiently.  Various truncation methods have been
developed during the last twenty years, for example the methods that
produce exponential convergence of observables as a function of
truncation~\cite{Horoi94,Papenbrock04,Andreozzi01}.  A very good example
of the methods that make SMCI calculations more efficient in terms of
computation resource usage is being used in the SMCI code {\it nathan}
and its cousin, the code {\it antoine}~\cite{Nowacki95,Caurier99}.
These codes use a novel Hamiltonian matrix compression method where the
SMCI proton-neutron Hamiltonian matrix elements are not stored before
a SMCI calculation, but constructed during each matrix-vector
operation of a Lanczos procedure using precalculated Reduced Density
Matrix (RDM) elements.  This innovation made it possible to make the
first unrestricted SMCI calculations in the pf-shell~\cite{Caurier99},
where other SMCI programs using older methods could not work without
truncations. For example in the nucleus $^{56}{\rm Ni}$ in the
pf-shell this method avoids the explicit storage of $10^{14}$ non-zero
Hamiltonian matrix elements.

The most basic SMCI method is the m-scheme method~\cite{Whitehead77}
that uses bare Slater determinants of spherically symmetric s.p. orbit
configurations as its many-body basis states. The basic problem with
the m-scheme SMCI is that the Slater determinant basis dimension is
maximal and therefore a lot of storage (from gigabytes to tens of
gigabytes) is needed for each calculated Lanczos basis vector in
large-scale calculations.  A common method to reduce the large matrix
dimensions of the m-scheme SMCI is to use the existing symmetries of
the nuclear Hamiltonian.  The {\it j-scheme} SMCI method uses angular
momentum projected many-body basis states, but does not have good
isospin, and is used in the SMCI code {\it nathan}.  Compared to the
m-scheme the j-scheme typically reduces the SMCI dimensions by two
orders of magnitude for low-spin states and less for high-spin states.
This property makes it most suitable for low-spin states, such as the
ground states of double-even nuclei.  The j-scheme
method~\cite{Caurier99,Nowacki95} of matrix-vector products is further
developed here to make it numerically more efficient and more suitable
for very large SMCI calculations.

% Put \label in argument of \section for cross-referencing
%\section{II. OVERVIEW OF THE METHOD} 

{\it Theoretical overview:} The j-scheme method forms the SMCI
many-body basis states using angular momentum coupled products of
proton and neutron many-body basis states. Lanczos vectors $\ket{\vec
  v_{JM}}$ are expanded as

\begin{eqnarray}
  \ket{\vec v_{JM}}&=&\sum_{j'_p\alpha'_p} \sum_{j'_n\alpha'_n}
  v^{j'_{n} j'_{p} J}_{\alpha'_n, \alpha'_p}
  \left[ \ket{j'_{n} \alpha'_n}\ket{j'_{p} \alpha'_p}\right]_{JM}\,.\\\nonumber
\end{eqnarray}

\noindent where $j_p$ and $j_n$ are proton and neutron total angular
momenta and the quantum numbers $\alpha_p$ and $\alpha_n$ sum over all
possible combinations of linearly independent states. Because of the
angular momentum coupled structure of basis states, the full
Hamiltonian matrix can be naturally divided into matrix blocks that
contain all allowed bra and ket particle configurations but where the
proton and neutron bra and ket angular momenta are kept fixed. One
such Hamiltonian matrix block will be concentrated on here. In
addition, only the proton-neutron part of the nuclear two-body Hamiltonian is
considered, whose treatment in the proton-neutron formalism is the
most time consuming part of a j-scheme SMCI calculation.

Using the angular momentum coupled states of (1), the matrix
elements of the nuclear proton-neutron Hamiltonian are sums of the products of proton
one-body RDM elements, neutron RDM elements, angular momentum
recoupling factors and transformed two-body interaction matrix elements:

% twocolumn version
%\begin{eqnarray}
%  \label{basic}
%  H_{i+j,i'+j'} &=& \bra{j_n \alpha_n j_p \alpha_p; J} \hat{H}^{(2)}_{\rm pn}
%  \ket{j'_n \alpha'_n j'_p \alpha'_p; J} \nonumber\\
%  &=& \sum_{aa'bb'\lambda}^{} \Gamma^\lambda
%  \redbra{j_p[\alpha_p]_j} [\cd{a}\ct{{a'}}]_\lambda
%  \redket{j_p'[\alpha_p']_{j'}}  \\
%  &\times&  \redbra{j_n[\alpha_n]_i} [\cd{b}\ct{{b'}}]_\lambda
%  \redket{j_n'[\alpha_n']_{i'}}F_\lambda(aa'bb') \,,\nonumber\\\nonumber
%\end{eqnarray}
\begin{eqnarray}
  \label{basic}
  H_{i+j,i'+j'} &=& \bra{j_n \alpha_n j_p \alpha_p; J} \hat{H}^{(2)}_{\rm pn}
  \ket{j'_n \alpha'_n j'_p \alpha'_p; J} \nonumber\\
  &=& \sum_{aa'bb'\lambda}^{} \Gamma^\lambda
  \redbra{j_p[\alpha_p]_j} [\cd{a}\ct{{a'}}]_\lambda
  \redket{j_p'[\alpha_p']_{j'}} \redbra{j_n[\alpha_n]_i} [\cd{b}\ct{{b'}}]_\lambda
  \redket{j_n'[\alpha_n']_{i'}}F_\lambda(aa'bb') \,,\nonumber\\\nonumber
\end{eqnarray}

\noindent as originally shown by French~\cite{French69}.
$F_{\lambda}(aa'bb')$ are particle-hole transformed two-body
interaction matrix elements and $\Gamma^\lambda$ is an angular momentum
recoupling factor (see~\cite{Caurier05} for more details).  Eq. (2)
shows the principal idea of~\cite{Caurier99,Nowacki95}: Create row and
column indices for proton and neutron RDM elements in such a way that
the indices of the full Hamiltonian matrix can be obtained just by
summing together the proton and neutron indices.  In this way a small
amount of RDM elements can generate a large number of Hamiltonian matrix
elements. The method of Eq. (2) will now be extended. To simplify the
subsequent formulae the $d_1\times d_1'$ neutron RDM is written as

\begin{equation}
  \redbra{j_n[\alpha_n]_j} [\cd{a}\ct{{a'}}]_\lambda
  \redket{j_n'[\alpha_n']_{j'}}\equiv A^{aa'\lambda}_{jj'}\,,
\end{equation}

\noindent where $j\in [1,d_1]$ and $j'\in[1,d'_1]$ and the $d_2\times d_2'$ proton RDM as

\begin{equation}
  \redbra{j_p[\alpha_p]_i} [\cd{b}\ct{{b'}}]_\lambda
  \redket{j_p'[\alpha_p']_{i'}}\equiv B^{bb'\lambda}_{ii'}\,,
\end{equation}

\noindent where $i\in [1,d_2]$ and $i'\in [1,d'_2]$.  Inside the RDMs
${\bf A}$ and ${\bf B}$ each set of quantum numbers $[\alpha_n]$ has a
unique row or column index $i$. The density operator indices label
sets of s.p. quantum numbers, $a=\left\lbrace
  n_a,l_a,j_a\right\rbrace$, $b=\left\lbrace
  n_b,l_b,j_b\right\rbrace$.  Since each density operator only
connects one bra particle configuration to one ket configuration, the
matrices ${\bf A}$ and ${\bf B}$ are sparse supermatrices that consist
of dense blocks.  Furthermore, a matrix ${\bf A}^{aa'\lambda}$ that
corresponds to a certain density operator $\bigl[ \cd{a}
\ct{a'}\bigr]_\lambda$ has only one dense matrix block on each
supermatrix block row.  

A Lanczos basis vector $\ket{{\vec v}_{JM}}$ and the result of a
Hamiltonian matrix-vector product $\ket{{\vec u}_{JM}}={\hat H}
\ket{{\vec v}_{JM}}$ can be ordered so that their basis state
amplitudes can be expressed in terms of matrices:

\begin{eqnarray}
  \ket{\vec v_{JM}} &=& \sum_{j'=1}^{d'_2}\sum_{i'=1}^{d'_1}
  v^{j'_{n} j'_{p} J}_{d_1(j'-1)+i'}
  \left[ \ket{j'_{n} [\alpha'_n]_{i'}}\ket{j'_{p}
      [\alpha'_p]_{j'}}\right]_{JM}\,,\nonumber\\
  &\equiv& \sum_{j'=1}^{d'_2}\sum_{i'=1}^{d'_1}
  V_{i'j'}  \left[ \ket{j'_{n} [\alpha'_n]_{i'}}\ket{j'_{p}
      [\alpha'_p]_{j'}}\right]_{JM}\,,\\\nonumber
\end{eqnarray}

\noindent and similarly for the vector $\ket{\vec u_{JM}}$ where the
constant quantum numbers have been omitted for simplicity. Using Eqs.
(2-5), the Hamiltonian matrix of Eq. (2) can be transformed to sums
over direct products of proton and neutron density matrices and the
matrix-vector product can be expressed as a sum of triple matrix
products:

\begin{eqnarray}
  U_{ji} &\equiv& u_{(i-1)d_1+j} =  H_{(i-1)d_1+j,(i'-1)d'_1+j'} v_{(i'-1)d'_1+j'} \nonumber\\
  &=& \sum_{\lambda aa'bb'} \Gamma^\lambda F_\lambda(aa'bb') A^{aa'\lambda}_{jj'} B^{bb'\lambda}_{ii'} v_{(i'-1)d'_1+j'} \nonumber\\
  &=& \sum_{\lambda bb'} \Bigl( \sum_{aa'} \Gamma^\lambda
  F_\lambda(aa'bb') A^{aa'\lambda}_{jj'} \Bigr) V_{j'i'} B^{bb'\lambda}_{ii'}\,. \\\nonumber
\end{eqnarray}

In this equation the indices $i'$ and $j'$ are implicitly summed over.
Alternatively, one may change the order of the matrix products and use the
equation

\begin{equation}
  U_{ji} = \sum_{\lambda aa'} A^{aa'\lambda}_{jj'}
  V_{j'i'}\Bigl( \sum_{bb'} \Gamma^\lambda F_\lambda(aa'bb')
    B^{bb'\lambda}_{ii'} \Bigr) \,,
\end{equation}

\noindent if it uses less sum and multiplication operations. Usually
for $N_v \neq Z_v$ nuclei the density matrices corresponding to
smaller number of valence particles/holes are smaller and should be used in the innermost
loop. Eqs. (6) and (7)  can still be further optimised. Because the same
density matrix ${\bf A}^{aa'\lambda}$ is used for more than one
Hamiltonian block, Eq. (6) can be converted to the form

\begin{equation}
  U_{ji} = \sum_{\lambda bb'} \Bigl( \sum_{aa'} F_\lambda(aa'bb')
    A^{aa'\lambda}_{jj'} \Bigr) \sum_{j'_n} \Gamma^\lambda V^{j'_n}_{j'i'}
    B^{bb'\lambda j'_n}_{ii'}
\end{equation}

\noindent where the quantum number $j'_n$ of a neutron many-body basis
state goes through all neutron angular momenta quantum numbers that
can be coupled with fixed proton ket angular momentum quantum number
$j'_p$ to a total angular momentum $J$. In this way multiple right-hand
side matrices ${\bf V}^{j'_n}$ can be multiplied with their
corresponding density matrices ${\bf B}^{bb'\lambda j'_n}$ and summed
together before performing the multiplications with the blocks of matrix ${\bf
  A}^{aa'\lambda}$.  Eq. (8) reduces to Eq. (6) for $J=0$
states, but reduces floating point operations for states with higher
angular momentum by approximately $30-50$\%. It is however quite complex to implement and
therefore has not yet been used in this work. {\it Note:} The form of Eqs.
(6-7) makes them ideal for separable interactions, such as the pairing
plus quadrupole interaction ${\hat H}^{(2)} = -G {\hat P}^\dg
{\hat P}^{} - \half\chi\sum_\mu {\hat Q}^\dg_\mu {\hat
  Q}^{}_\mu$ or the center-of-mass interaction~\cite{Lawson79},
because the resulting separability of the two-body interaction matrix
elements, $F_\lambda(aa'bb')=f_\lambda(aa')f_\lambda(bb')$, allows the
full matrices $\bf A$ and $\bf B$ in Eqs. (6-8) to be used totally independently of each
other.

%\section{III. BENCHMARKS AND COMPARISONS} 

{\it Discussion:} The results of unrestricted benchmark calculations
made in the pf-shell model space for nuclei from $^{44}{\rm Ti}$ to
$^{56}{\rm Ni}$ using an implementation of this method in the SMCI
code {\it eicode}~\cite{Toivanen04,Kortelainen06} are presented in
Fig. 1.  All results are calculated using one AMD Opteron $2.4$ GHz
processor.  It can be seen that the number of mathematical operations
per one Hamiltonian matrix-vector product scales as $d^{1.46}$ for
angular momenta $J=0,2$ where $d$ is the SMCI matrix dimension. The
number of mathematical operations for higher angular momenta (which
have exactly the same scaling) have not been included for clarity. In
this work the best implementation of the original j-scheme matrix
vector product method of~\cite{Caurier99,Nowacki95} scales as $d^{1.82}$ as a
function of basis dimension (Fig. 1, dashed line). Whereas
calculations for all angular momentum values scale exactly similarly
in the case of the old method, the new method has slightly different
constant multiplicative factors that depend on angular momentum and
the isospin z-component. The dependence on angular momentum is roughly
$2J+1$ and can be removed to a large extent by using Eq. (8) for
matrix-vector products instead of Eqs. (6-7). Both the old method and
the method of Eqs. (6-7) use the largest number of floating point
operations for nuclei with $N_v=Z_v$, and therefore the wavefunctions of
these nuclei are the most time consuming ones to calculate.

\begin{figure}
  \includegraphics[angle=-90,width=8.6cm]{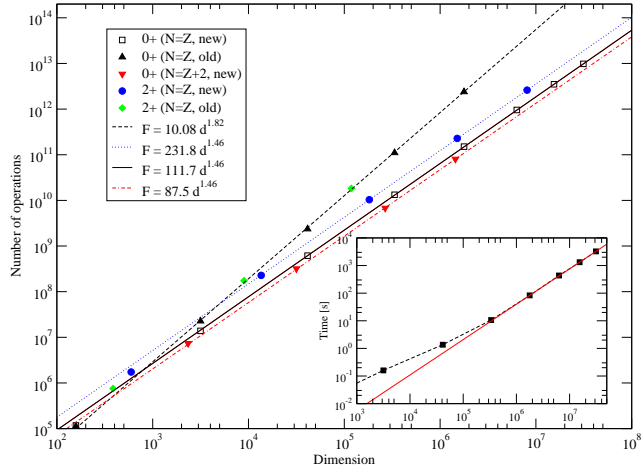}%{vbr_scaling.ps}
  \caption{\label{} (Color online) Number of mathematical operations needed for one
    matrix-vector product as a function of basis dimension for various
    $N_v=Z_v$ and $N_v=Z_v+2$ pf-shell nuclei from $^{44}{\rm Ti}$ to
    $^{60}{\rm Zn}$.  The lines are fit functions, shown in the legend
    box.  Inset: Actual time per one matrix vector operation for $0^+$
    states of $N_v=Z_v$ pf-shell nuclei using the new method. The solid
    line is a fitted function $F=8.28\cdot 10^{-7} d^{1.28}$.}
\end{figure}

The more favourable scaling of floating point operations in this
modified method reduces them significantly for very large
calculations. In the case of $0^+$ states of $^{56}{\rm Ni}$ the
reduction of floating point operations is $35$-fold.  Considering the
actual matrix-vector product times, the time plot in Fig. 1 inset
shows two different regimes.  In the low-dimensional cases matrix
products are very fast, initialisation overheads dominate the
calculation time, and therefore only the scaling of matrix-vector
operation times in the higher-dimensional regime from $^{50}{\rm Mn}$
onwards are of interest. In this regime the matrix-vector operation
{\it times} scale as $n^{1.28}$. This scaling can be compared against the
$n^{1.1}$ scaling of the m-scheme code {\it antoine}, shown on page 445
of~\cite{Caurier05} where the time for $15$ Lanczos iterations for
$0^+$ states of the pf-shell nucleus $^{52}{\rm Fe}$ is roughly $10^4$
seconds and roughly $10^5$ seconds for $^{56}{\rm Ni}$ (extrapolated),
giving average matrix-vector multiplication times of $670$ and $6700$
seconds using a modern microprocessor comparable to the one used in
this work.  For the same two calculations the new j-scheme method that
uses Eqs. (6) and (7) results with one Hamiltonian matrix-vector
multiplication taking respectively $83$ and $1317$s.  
The code {\it antoine} is therefore five times slower than the new
j-scheme method of Eqs. (6-7) for $^{56}{\rm Ni}$. Compared to the original j-scheme
method of ~\cite{Caurier99,Nowacki95}, used in {\it eicode},
the new method shortens the calculation time $75$-fold for this
nucleus. 

\begin{figure}
  \includegraphics[angle=-90,width=8.6cm]{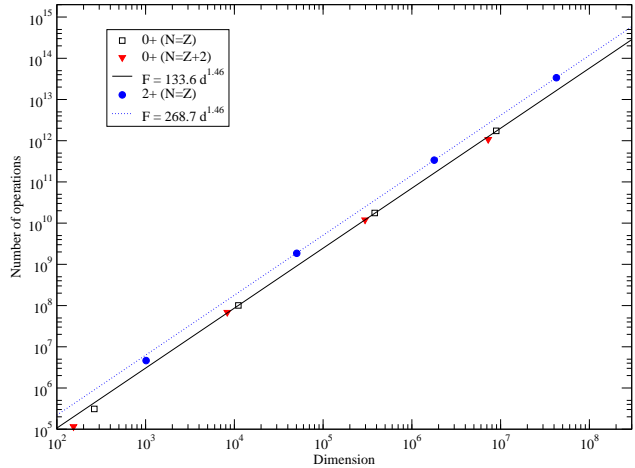}
  \caption{\label{} (Color online) Same as Fig. 1, but for unrestricted calculations
    for $N=Z$ and $N=Z+2$ nuclei in the $sdg_{7/2}h_{11/2}$
    model space using the method of Eqs. (6-7).}
\end{figure}

The reason why the new method is faster than what the reduction of the
number of mathematical operations alone in Fig. 1 suggests, is that the
matrix products of Eqs. (6-7) can be implemented very efficiently on a
modern computer using fast matrix product kernel
routines~\cite{Goto1}.  For large SMCI calculations, where the RDM
blocks have large dimensions, the new method can use up to $70\%$ of a
processor's theoretical peak floating point capacity in the pf-shell
model space and even slightly more in larger s.p. model spaces.  In
contrast, the efficiency of the best implementation of the original
j-scheme method of~\cite{Caurier99} in the code {\it eicode} is always
low, roughly $10-15\%$ of the processor's peak capacity. Looking at
the calculation time plot of~\cite{Caurier05} it is suspected that
this is also the case for the codes {\it antoine} and {\it nathan}.
The reasons for this inefficiency are the explicit calculation of
matrix indices and the explicit construction of {\it every} Hamiltonian
matrix element, which are avoided in the new method.

{\it Conclusions:} It has been shown that the numerical effort needed
to use the optimised j-scheme matrix-vector products of Eqs. (6-7)
scales more favourably than the old j-scheme method, used in the SMCI
codes {\it nathan} and {\it eicode}, as a function of basis dimension,
making the new method up to two orders of magnitude faster in the
pf-shell. The same speedup and scaling of floating point operations
also happens in the ${\rm sd}{\rm g}_{7/2}{\rm h}_{11/2}$ model space
used for nuclei above $^{100}{\rm Sn}$.  This can be seen from Fig.
2, where the scaling of floating point operations has been plotted for
various $N_v,Z_v=3-8$ nuclei in this model space. The increased
efficiency makes the method competitive against the m-scheme SMCI
method of~\cite{Caurier99,Nowacki95} that is used in most modern
m-scheme SMCI codes.

\begin{table} %[H] add [H] placement to break table across pages
  \caption{\label{}M-scheme and j-scheme dimensions, number of
    floating point operations and estimated
    computation time with one CPU for various unrestricted SMCI
    calculations for the $0^+$ states of nuclei in the
    $sdg_{7/2}h_{11/2}$ model space. $N_v$ and $Z_v$ are the number of
    valence neutrons and protons.}
  \begin{ruledtabular}
    \begin{tabular}{ccccc}
      $(N_v,Z_v)$ & Dim (m) & Dim (j) & $F$ & $T_1$[s] \\ 
      \hline
      $(6,4)$     & $8.56\cdot 10^{8}$  & $7.24\cdot 10^{6}$  & $1.38\cdot 10^{12}$ & $550$ \\ 
      $(5,5)$     & $1.06\cdot 10^{9}$  & $8.92\cdot 10^{6}$  & $1.73\cdot 10^{12}$ & $870$ \\ 
      $(7,5)$     & $1.65\cdot 10^{10}$ & $1.21\cdot 10^{8}$  & $8.45\cdot 10^{13}$ & $2.9\cdot 10^4$ \\ 
      $(6,6)$     & $2.00\cdot 10^{10}$ & $1.45\cdot 10^{8}$  & $1.02\cdot 10^{14}$ & $3.4\cdot 10^4$ \\ 
      $(8,6)$     & $2.20\cdot 10^{11}$ & $1.43\cdot 10^{9}$  & $3.11\cdot 10^{15}$ & $1.03\cdot 10^6$ \\ 
      $(8,8)$     & $2.43\cdot 10^{12}$ & $1.50\cdot 10^{9}$  & $3.08\cdot 10^{15}$ & $1.02\cdot 10^6$ \\ 
    \end{tabular}
  \end{ruledtabular}
\end{table}

With the j-scheme matrix vector product method shown here it is
possible to make unrestricted SMCI calculations for all nuclei in the
${\rm pf}$ model space and for most nuclei in the ${\rm pf}_{5/2}{\rm
  g}_{9/2}$ model space using modest computing resources. For more
complex model spaces more computational power will be necessary, but
calculations still stay tractable, since j-scheme dimensions of the
order of $2\cdot 10^9$ can be handled easily. Table 1 shows estimated
matrix-vector product times for various nuclei in the
$sdg_{7/2}h_{11/2}$ model space for up to eight valence protons and
neutron holes. For up to $10$ valence particles+holes the method of
Eqs.  (6-7) can be used without large computational resources. The
calculations with more than $10$ valence particles need larger
computational resources.  Using the original j-scheme method
of~\cite{Caurier99,Nowacki95}, at least two orders of magnitude more
resources would have to be used to obtain results equally quickly for
large-scale calculations. The m-scheme SMCI methods also encounter
their dimensionality problem with these nuclei.

Dimensions of the same order of magnitude as in Table 1 will also be
encountered in the $pf_{5/2}g_{7/2}d_{5/2}$ model space that is used
for the description of deformed nuclei in the $N\approx Z = 40$ region
and with the No-Core Shell-Model (NCSM)~\cite{Navratil06} used for
the {\it ab initio} description of the structure of light nuclei.  The
method described here makes it possible to do j-scheme SMCI
calculations up to a dimension of $10^{10}$ with current computing
hardware and is a step towards unrestricted calculations in the
aforementioned model spaces.  A further increase of SMCI calculation
dimensions can be sought for by combining this method with the
advanced truncation methods that show exponential convergence, such as
the methods of Refs.~\cite{Horoi94,Andreozzi01,Papenbrock04}.

\end{document}